\documentclass[11pt,twoside]{article}

\usepackage{asp2014}

\aspSuppressVolSlug
\resetcounters

\begin{document}

\title{Separating detection and catalog production}

\author{Mohammad~Akhlaghi$^1$
\affil{$^1$CRAL, Observatoire de Lyon, CNRS, Universite Lyon 1, 9 avenue Ch. Andre, 69561 Saint Genis-Laval Cedex, France; \email{mohammad.akhlaghi@univ-lyon1.fr}}}

\paperauthor{Mohammad~Akhlaghi}{mohammad.akhlaghi@univ-lyon1.fr}{orcid.org/0000-0003-1710-6613}{CRAL}{Observatoire de Lyon}{Saint Genis Laval}{Auvergne-Rhône-Alpes}{69561}{France}

\begin{abstract}
In the coming era of massive surveys (e.g. LSST, SKA), the role of the
database designers and the algorithms they choose to adopt becomes the
decisive factor in scientific progress. Systems that allow/encourage
users/scientists to be more creative with the reduction/analysis
algorithms can greatly enhance scientific productivity. The
separation/modularity of the detection processes and catalog
production is one proposal for achieving `Reduction/analysis
algorithms for large databases and vice versa'. With the new
noise-based detection paradigm, non-parametric detection is now
possible for astronomical objects to very low surface brightness
limits. In our implementation, one software (\textsf{NoiseChisel}) is
in charge of detection and another (\textsf{MakeCatalog}) is in charge
of catalog production. This modularity has many advantages for
pipeline developers, and more importantly, it empowers scientific
curiosity and creativity.
\end{abstract}

\section{Introduction}

At the lowest level, each datum, or pixel in an image, only has two
properties: its value and its position relative to the
others. Therefore, raw data, or an image in the case of this paper, is
not directly usable for scientific analysis. For that, we
\emph{reduce} the low-level raw data set (image) into a more formal and
higher-level structure like catalogs. In the most basic terms, the
conversion/reduction from an image to a catalog can be described as:

\begin{enumerate}
\item Detection: identify the pixels associated to each target.
\item Measurement: calculate various properties on similarly labeled
  pixels from detection. For example, the magnitude of an astronomical
  object, which can be the sum of the pixel values of each label. The
  center or position of the object can be the average position of the
  pixels associated with it, weighted by the (normalized) pixel
  values.
\end{enumerate}

Detection (step 1) hence defines the number of objects, or rows or
records and their pixel footprints. Afterwards step 2 can be run
separately for each desired property (column, or field) in the final
catalog. The resulting rows (from step 1) and columns (from step 2)
create a catalog which can then be used for higher-level processing
(for example estimating the stellar mass using magnitudes measured on
multiple filters).

When the targets have a sharp edge (for example cells in
medical/biological imaging that have a clear membrane separating them
from the background), a threshold that is sufficiently lower than the
edge value and sufficiently above the noise level, will be able to
clearly detect all the objects and separate their pixels from the
noise. This enables the creation of catalogs with only one pass
through the steps above. Because the threshold is defined to avoid
noise, this method can be classified as signal-based detection.

However, most astronomical targets, do not have a strong edge, for
example galaxies, nebulae, stars (or the PSF), and comets. The signal
of nearly all astronomical objects sinks into the noise very
gradually, see Figure 1b of \cite{2015ApJS..220....1A} for an
example. Any threshold that is defined \emph{to avoid} noise will
inevitably miss a significant fraction of the object's flux or
structure. The solution until now has been to make multiple passes
through the steps above: in the first pass, the regions above the
threshold are identified and first and second moment measurements
(center, and elliptical parameters) are made for them along with other
measurements. These measurements are then used to model the brighter
parts in order to extrapolate the model to regions below the
threshold.

Some traditional applications of this multi-pass approach are the
\citet{1976ApJ...209L...1P} and \citet{1980ApJS...43..305K}
methods. The successful application of this process depends on the
object having a single and simple elliptical (or model-able) profile,
which is idealistic.  Various measurements (for example the center,
ellipticity, Sky value, radial distribution of flux and etc) are also
necessary to make these modelings. The iterative nature of catalog
production in the signal-based detection paradigm, thus creates
systematic biases and adds complexity. Technically, it makes catalog
production a very computationally expensive process that can decrease
creativity.

\begin{figure}[t]
  \includegraphics[trim=0cm 1.7cm 0cm 1.8cm, clip,
    width=\linewidth]{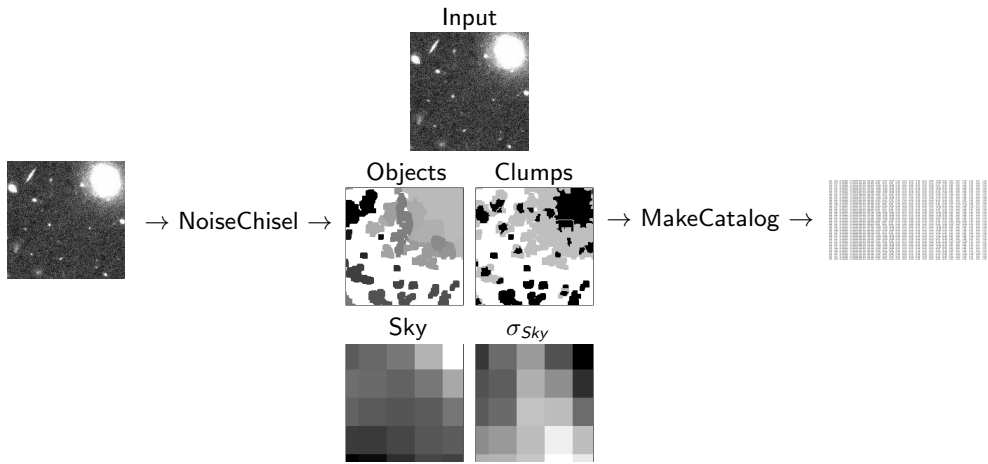}
  \caption{\label{O8-1-f1} \textsf{NoiseChisel} and
    \textsf{MakeCatalog} inputs and outputs. \textsf{NoiseChisel}
    produces a 5 extension FITS file which contains the input image,
    the clumps and objects labeled images and the Sky and Sky standard
    deviations over a mesh grid. These images are then fed into
    \textsf{MakeCatalog} to generate a catalog. This modularization
    enables the users to mix different \textsf{MakeCatalog} inputs, or
    define their own labeled images, or noise
    properties. \textsf{MakeCatalog} is also a simple program, so
    adding new columns internally is very easy.}
\end{figure}

A new noise-based detection method was introduced in
\citet{2015ApJS..220....1A} and also in the 25th ADASS. In this
method, the threshold is below the Sky value and not intended to avoid
noise, but embrace and benefit from it. Signal is separated from the
noise by exploiting the 2D contiguity of the pixels that contain
signal. It is thus able to detect very diffuse structure of any shape
without any parametric modeling. Therefore with this approach, it is
no possible to generate a scientifically useful catalog with only one
pass of the steps above. \textsf{NoiseChisel} is the name of our
software implementation and is distributed as part of the GNU
Astronomy Utilities (Gnuastro). To further emphasize the distinction
between the two steps, its outputs are labeled images and noise
properties (the Sky and Sky standard deviation), see Figure
\ref{O8-1-f1}.

The input image, the labeled image(s) and the noise properties (Sky
value and its standard deviation) are then all fed into another tool
(\textsf{MakeCatalog}) to generate a catalog. Separating catalog
production from detection (with labeled and noise images as the
intermediate state) allows a new degree of freedom to the scientists:
the ability to access the pixels of each object, while also improving
developer experience because of adherence to the Unix philosophy: 1) Do
one thing and do it well. To do a new job, build afresh rather than
complicate old programs by adding new "features".  2) Expect the
output of every program to become the input to another program. For
example, to add new columns to the catalog, only one function and
several variables have to be added in the source of the small
\textsf{MakeCatalog}
program\footnote{\url{https://www.gnu.org/s/gnuastro/manual/html_node/Adding-new-columns-to-MakeCatalog.html}}
instead of having to delve with the much larger and complicated
detection program (\textsf{NoiseChisel}).

Technically, a labeled image (the ``Objects'' and ``Clumps'' images of
Figure \ref{O8-1-f1}) only has integer values (for example 0 for the
sky regions and a positive integer for each object in the table). So
it can be highly compressed compared to the original image, or a large
catalog with many columns (and easily transferred over the network or
stored within a database). Furthermore, the labeled image can be
broken up into small crops, each containing one object's pixels (with
WCS information to easily match to the correct part of the
image). Since this box only belongs to one detection, it is possible
to define one bit for every pixel and improve the compression ratio
even further. A binary image like this can be used when there is no
deblending. When sources blend, an 8-bit per pixel value can be
assigned enabling 255 layers to assign a weight to the pixel
values. Since the largest number of objects in deep astronomical
images are small and faint galaxies, the majority of object labels can
be stored with one bit per pixel.

With this modularity, the expensive detection process can be run with
various input parameters by the pipeline developers, enabling pipeline
users to choose which ever set of parameters best suites their
science. For example, the completeness and purity of a detection
algorithm are anti-correlated: allowing lower purity (more
contamination by false detections) improves completeness (detecting
true detections). When the science case involves detection in multiple
colors (images), lower purity can be corrected, because a false
detection will not be present in multiple images. For example, in the
definition of Lyman break dropout galaxies, we expect detections in
all filters red-ward of the break. However, since the Sky value is the
average of undetected pixels, one problem to doing this generally is
that a large number of false detections will underestimate the Sky
value (by systematically removing localized noise peaks). So the Sky
value and its standard deviation can be taken from other detection
runs with a more reasonable purity. So in this example, high-redshift
studies like this can greatly benefit from the improved completeness
that is provided by this flexibility and modularity.

In this modular approach, photometry over an ellipse/aperture, or even
Kron or Petrosian photometry is also possible. To do that,
\textsf{MakeCatalog} can be used to get elliptical and other
parameters from the raw \textsf{NoiseChisel}
detections. Elliptical/aperture labeled regions can then be created
based on the derived properties in the initial catalog. In Gnuastro,
\textsf{MakeProfiles} can make such labeled
ellipses/apertures. \textsf{MakeCatalog} can then be told to use the
elliptical labeled image as the ``Objects'' input image to create a
new catalog\footnote{If the ``Objects'' input to \textsf{MakeCatalog}
  (see Figure \ref{O8-1-f1}) was not created by \textsf{NoiseChisel},
  no ``Clumps'' image will be used/necessary and no clumps catalog
  will be created.}. Alternatively, if aperture photometry is
necessary on a-priori known positions, \textsf{NoiseChisel} detections
can be ignored and only its Sky and Sky standard deviation outputs can
be used. The apertures (in any circular or elliptical shape) can be
created as a segmentation map with \textsf{MakeProfiles} which can be
fed into \textsf{MakeCatalog}. In a large database, the Sky and Sky
standard deviation images are internally stored, so the user just has
to define their labeled images or segmentation maps (which are highly
compressed and easy to upload as discussed above).

Another application of this modularity is matched photometry, when the
same pixels need to be used in multiple filters/images to obtain
colors. In this scenario, the same segmentation map can be used with
multiple filter images (and their Sky and Sky Standard deviation
images) to generate such a catalog. With this, users can easily get
multi-color catalogs from different surveys that give images in
different filters. The labeled regions can be taken from any survey
and fed into another survey. The user can account for varying
resolution and PSF on the labeled images by warping and convolving to
the other survey's resolution.

Each pixel ultimately has just two values (its position and its
value), in task 14244 we plan to add a feature to \textsf{MakeCatalog}
that will allow the users to define their own columns with a very
simple syntax. In this way the user can directly access the pixels to
generate their own high-level catalog best suited for their particular
science, without having to internally edit/modify the code and rebuild
it. It is also possible to add dynamic loaded libraries (plug-ins) so
developers can compile their own column creation libraries and load
them into \textsf{MakeCatalog}, both on survey servers or individually
to share their work without bloating the root \textsf{MakeCatalog}
program.

Gnuastro (\url{https://www.gnu.org/s/gnuastro/}) is the parent
software project to \textsf{NoiseChisel}, \textsf{MakeCatalog}, and
\textsf{MakeProfiles} and many other programs and libraries for
astronomical data analysis and manipulation. It has very few
dependencies and is portable to all Unix-like operating systems. All
its utilities are defined based on the Unix philosophy with maximal
modularity, simplicity and efficiency. It is heavily documented with a
complete manual and also comments. Therefore similar to Unix-like
operating systems it can be run on small home computers, or large
survey databases.

\acknowledgements I gratefully acknowledge support from the ERC
advanced grant 339659-MUSICOS.

\bibliography{O8-1}  

\end{document}